\documentclass[twoside]{article}
\usepackage{fleqn,espcrc2}

\newcommand{\AmS}{{\protect\the\textfont2
  A\kern-.1667em\lower.5ex\hbox{M}\kern-.125emS}}

\title{Holography of charges in gauge theories \hfill LPT-ENS/01-21 }

\author{B.L. Julia\address{Laboratoire de Physique th\' eorique \\
de l'Ecole Normale Sup\' erieure,\\
  24 rue Lhomond 75005 Paris (France)}
        \thanks{UMR  8549 du CNRS, research supported in part  by EU TMR 
contract ERBFMRXCT96-0012. Proceedings of the D.V. Volkov Memorial conference 
Kharkov 2000 ed. D. Sorokin et al.}}

\def\ex{\mbox{\boldmath$d$}}
\def\we{\mbox{\footnotesize \boldmath$\wedge$}}
\def\p{\partial}
\def\d{\delta}

\begin{document}

\begin{abstract}
In this short review we compare the rigid Noether charges to  
topological gauge charges. One important extension is that  one should consider 
each boundary component of spacetime independently. The argument 
that relates bulk charges to 
surface terms can be adapted to the perfect fluid situation where one can 
recognise the 
helicity and enstrophies as Noether charges. More generally a forcing procedure 
that increases for instance any Noether charge is demonstrated. In the gauge 
theory situation, the key idea can
be summarized by one sentence: ``go to infinity and stay there''.
A new variational 
formulation of Einstein's gravity is given that allows for local GL(D,R) 
invariance. The a priori indeterminacy of the Noether charges is emphasized 
and a covariant ansatz due to S. Silva for the surface charges of 
gauge theories is analysed, it replaces the (non-covariant) Regge-Teitelboim 
procedure. 
\vspace{1pc}
\end{abstract}

\maketitle

\section{Introduction}

Between 1918 and 1961 \cite{EN,ADM} there was no clearly defined total 
charge in gauge theories and in particular in General Relativity. The 
few  physicists who have read Noether's paper will recall what we might 
call Hilbert's disaster; namely the property of any global current in a 
gauge theory to be a total divergence. In those days the asymptotic 
structure of spacetime had not been formalised not to speak of 
the topological charges.
We shall explain this statement and recall that in the Maxwell case
it is so obvious that we do not pay attention, what happens is that 
the Noether current is on-shell equal to a topological current (by 
Gauss' law). 

The mathematical reasoning can be adapted to the situation of time independent 
gauge invariance as it is encountered in perfect fluids. There the dynamical 
object is a map from lagrangian coordinate volume to the laboratory vessel, 
even for a compressible fluid the adiabaticity and isentropy conditions ensure 
that all Lagrangian volume elements are indistinguishable and hence guarantee 
time-independent diffeomorphism  invariance, actually volume preserving ones.
We are in a situation where gauge invariance is realised without  gauge fields,
no metric is available (general relativistic fluids would be presented with a 
metric in laboratory coordinates). For Gedanken vessels without boundary there 
still can be helicity or enstrophies, the conserved charges in 3 (resp.2) 
space dimensions. The helicity has been analysed by H. K. Moffatt see \cite{AK} and the reference to J. J. Moreau. 
What may be surprising for a general relativist 
is that we do not need to use a boundary component nor special asymptotics, what
replaces that is a field dependent one-parameter subgroup  of the time 
independent volume preserving diffeomorphisms, its existence is in turn a
consequence of the gauge symmetry namely the Helmholtz-Kelvin theorem \cite{RS}.
The topological looking character of helicity, namely its interpretation as 
an average linking number of vortex lines, clashes clearly with our Noetherian 
interpretation especially in view of a variant of the following result. 

In passing we mention a general lemma obtained in \cite{JSI} about the 
forcing of Noether charges under rather general conditions, the assumption 
of velocity independent transformation can be relaxed for fluids. The idea 
is that contrary to the use of a boost in relativistic or galilean physics 
which 
requires 3 symmetries, for instance time  and space (x) translations and boosts 
mixing x and t coordinates and their charges, 
we actually need only one symmetry to be able to generate a solution
with nonzero-momentum from a static solution. The price to be paid is that one 
learns only how to modify initial conditions, yet although the resolution of 
the dynamical problem is left open, it is not needed and
one knows that the charge has increased 
after the appropriate kick.
 
The transmutation of bulk charges into boundary charges in gauge theories 
carries difficulties and advantages. The main misconception of most textbooks 
is that the so-called improvement terms of the currents is under control, it 
actually requires serious work. On the other hand the possibility to work at 
infinity (more precisely at one component of infinity) allows for singularities 
inside and the discussion should be entirely focussed away
from those (but for the 
variational principle which strictly speaking requires handling of all boundary components!).

Special cases like global (bulk) Killing vectors permit a standard Noetherian
treatment when they exist. Subtle global questions arise as well for locally 
asymptotically flat solutions for instance; there one must distinguish the local
measurement of the asymptotic field and the total charge value  assuming 
spherical symmetry, both may obey a different relationship than in the classical
case.  Finally p-form gauge invariances follow a similar pattern \cite{BJ}.    

\section{Surface terms and arbitrary ``improvements''}
It is easy to prove that when a rigid symmetry is actually a subgroup of a 
gauge symmetry its Noether current is a total divergence. The proof \cite{PB}
see the works of Bergmann's school relies on locality and an expansion in
derivatives of the symmetry parameter at a point: if they are all independent 
the result follows. For instance in the case of second order equations of motion
one obtains 
\begin{equation}
J_\xi := \xi . J + \ex \xi . \we U
\end{equation}
which is valid for all $\xi$ displacements. Henceforth $\ex J_{\xi}=0$ 
implies $J = \ex U$.
$U$ is traditionally called a superpotential.

This had been also discussed (actually independently) in \cite{BJ}. There 
gauge invariance was slightly enlarged by allowing gauge parameters that are 
closed 
but not exact, then the generalised Noether theorem for closed one-forms is 
precisely this total divergence formula. The extension  to arbitrary closed 
p-forms 
is straightforward, and the closed 0-forms are precisely the constant parameters
of the original Noether theorem.

Here comes an obvious corollary namely that the currents which are well known 
to be defined up to a total divergence when one 
modifies the action by a surface term that changes the boundary 
variation, are now totally arbitrary in a gauge context 
unless restricted by some extra requirement \cite{JSI}.   

Let us assume that the variation of the fields depend on the gauge parameter and
its derivative, then U depends up to a surface contribution only on the 
derivative part of the variation. When gauge invariance is implemented without 
gauge field it happens that U vanishes and hence the current as well
(2d conformal theories, kappa symmetry...). In the fluid case the invariance is 
only under volume preserving transformations the current is given by the sole
Lagrange multiplier for this constraint and the ``conservation'' of vorticity
obtains after taking a curl that kills the latter.

\section{Fluid flow invariants}

The action of isentropic fluids is given by \cite{RS}
\begin{equation}
L = \frac{1}{2} \left( \frac{\p x^i}{\p \tau} \right)^2 - e \left( \det{\frac
{\p x^i}{\p a^a}},\ s(a^a) \right) - \Phi (x^i)
\end{equation}
 The basic fields of our theory are the fluid-particle Eulerian 
coordinates $x^i (a^a,\tau)$. The cells of fluid are labelled by the $a^a$ at a 
given time $\tau$. The $i,j,k...$ indices will be used for the laboratory space 
(called $x$-space) whereas the $a,b,c...$ will be for the internal label space 
(called $a$-space) both with the  same dimension $D$. 
The labelling follows the fluid 
particles along the dynamics. The labels are the Lagrangian coordinates.
 $\Phi (x^i)$ is the potential of some external force. $e$ is just the 
specific internal energy, a given thermodynamic function of
 $\det{\frac{\p x^i}{\p a^a}}$ and of the specific entropy $s$. The important 
hypothesis here is that the entropy $s$  depends neither on the labels nor 
on the 
time $\tau$, these are respectively homentropy and 
adiabaticity or isentropy conditions.

In a gauge theory there are infinitely many one parameter subgroups of the 
gauge group, each one of them has its conserved Noether charge, they are in 
general useless as they are gauge dependent. What saves our day is that for a 
very specific field dependent transformation the current is in fact physical.
We can choose diffeomeorphisms that preserve the volume and are time 
independent by taking them as constant multiples of the vorticity. The 
resulting charge in 3 dimensions is nothing but the celebrated helicity that has important applications in turbulence and in  particular in 
magnetohydrodynamics \cite{CRAS}. 
The Moreau Moffatt helicity reads in Eulerian coordinates
as the Hopf index for the velocity (viewed as a potential one form):
\begin{equation}
H_M=\int v\wedge dv.
\end{equation}

Let us now present our forcing lemma which might be useful at least for 
numerical simulations as it will involve in the present situation  the 
Lagrangian coordinates. The latter are notoriously important in turbulence and 
regrettably hard to access experimentally. So let us assume that a Noether 
rigid invariance does not involve the time derivatives of the coordinates of a 
Lagrangian system. The fluid case requires more general hypotheses and we refer to the paper \cite{JSI} for details.

Lemma: given a global or rigid Noether symmetry and 
 the associated charge  given, with $x$ representing the coordinates or
field variables,
by $\d x=\xi (x)$ and $Q= \int J_\xi$ , 
then the change of $Q$ under the impulsive forcing at some time t given with
u the corresponding velocities by
\begin{equation}
\d x=0 \, , \, \d u=\xi (x) 
\end{equation}
\noindent is precisely equal to 
$$\d Q = (\p^2 L/\p u \p u). \xi .\xi$$
This is a positive quantity in view of the positivity of the acceptable 
kinetic terms.

Clearly  a random change in the initial conditions will change the 
value of the charge as it is not in general a symmetry transformation, the 
interesting fact is that our specific kick is in a sense optimised to increase 
the charge. Let us mention also that in all odd space dimensions one finds one 
conserved charge (at least?)  and in the even case an infinite number of them.

\section{Local GL(d,R) formulation of General Relativity} 

Let us now return to our main discussion of relativistic gauge invariant 
theories. It turns out to be much easier to use first order formulations to 
analyse the conserved currents. Surprisingly a new formulation of Einstein's 
gravity remained to be discovered. It lies above both the Palatini formalism 
in which metric and torsion free connection are taken as the independent 
variables and the Cartan-Weyl tetrad formulation where local Lorentz 
invariance is implemented through the introduction of the moving frames and the compensating non-propagating Lorentz connection. In this new variational 
principle the metric, the moving frames
$\theta^a$ and the linear connections $\omega^a_{\ b}$ are all 
independent. One obtains the previous formulations by gauge fixing and partial 
extremisation. Let us mention one interesting feature namely a
projective invariance under modifications of the linear connection by an 
arbitrary Weyl type diagonal part
\begin{equation}
\d_\kappa \omega^a_{\ b} = \kappa \d^a_{\ b}
\end{equation}
\begin{equation}
\d_\kappa \theta^a =\d_\kappa g^{ab} = 0
\end{equation}
where $\kappa$ is an arbitrary one-form.

The charge corresponding to a careful treatment of boundary terms turns out to 
be the KBL energy \cite{KBL,JSII} in the case of asymptotically flat boundary 
conditions. 

A general consequence of symmetries is the existence of
relations between equations of motion.
In the rigid Noetherian case the conservation of the current is such a 
differential relation. In the gauge case the Noether identities are algebraic 
relations expressing the fact that some variations vanish identically. 
The interesting situation arises here \cite{JSI}, see references 
therein and in particular \cite{DVM}, 
that all the equations of motion are in fact consequences of the 
symmetry in the sense that they follow from the conservation laws, this is 
typical of unique geometric Lagrangians (at a given order).

\section{Holography}

In  the rigid symmetry situation where the Noether procedure does provide a 
simple and general construction of currents up to topological terms, the 
conservation of charges follows from an extra hypothesis namely that there is
no leak of charge at infinity, the corresponding flux should vanish in the
use of Stokes theorem between two equal time surfaces. 

In the gauge case where the charge is given by a flux at infinity at a given 
time, again one must assume that there is no leak but the proof uses Stokes 
theorem at infinity only. It may be useful to emphasize that in that context
not only the current is ill defined, pseudo-energy momentum tensors proliferate
out of control, but the total bulk charge is also ill-defined. At least this is 
the case  whenever there is a singularity even when it is hidden behind an 
horizon. It is only recently and in special cases that a mass at the horizon 
has been constructed. Clearly one needs the analog of the asymptotic flatness
condition there or some well posed boundary condition, this has been realised 
only for the so-called isolated case \cite{ACK} up to now. 
The bulk charge only exists in fine as the sum of all the boundary 
contributions.

Let us now recall the Regge Teitelboim hypothesis \cite{RT}.
First of all a variational principle requires a well defined choice of boundary 
conditions for each component of the boundary. The symmetries to be discussed 
are those that preserve these asymptotic constraints. For example in the 
asymptotically flat situation the metric tends in a well defined way to the 
flat metric and the remaining symmetry at infinity is the rigid Poincar\'e 
group. In the case of the Einstein-Hilbert scalar action it is well known that 
one must modify the action by a surface term involving the second fundamental 
form, alternatively there is a non scalar action that only involves first 
derivatives of the metric through its quadratic dependence on the connection.

The gauge constraints due to reparametrisation invariance can be written as   

\begin{equation}
G(\xi,t) = \int \xi^A G_A dx
\end{equation}
where the gauge parameter $\xi$ tends to zero at infinity for proper gauge 
transformations. Under an arbitrary variation one may following \cite{RT}
require that the variation of $G$ be a bulk integral of an expression 
proportional to the local variations of the fields, physically it means that 
the degrees of freedom are in the bulk. Now if one applies this principle to the
conserved charges ie the same expressions $G$ for parameters that are nonzero 
at infinity one finds that they must be modified by surface terms so as to 
restore the bulk feature of the variation, in fact the evaluation of charges
would have been zero otherwise as the constraints vanish onshell. It is a little
bit paradoxical to restore bulkiness through a surface term, this is holography
at its best.
What has been shown in \cite{S} is that the resulting variational equation for 
the superpotential $U$  can be formulated in a covariant fashion. The 
integrability condition has been recently discussed in \cite{WZ} see also the 
many references to R. Wald's earlier work in there. A different approach can be found in \cite{AT}. In \cite{JSIII} we show that the construction of the 
symplectic form actually resembles the current discussion and compare various 
approaches.

Numerous applications confirm the above hypothesis \cite{HJS,JSII}. Typically 
naive approaches are prone to mistakes by factors of two as there is no 
systematics at all.

In conclusion we believe there are important open questions left open. The 
canonical treatment of null infinity is not systematic, horizons are still 
problematic in general, extended objects bring many new questions to our mind,
all of them are  rather urgent.

\end{document}